\documentclass[onecolumn]{aa} 

\usepackage[varg]{txfonts}
\usepackage{graphicx,subfigure,color,amsmath,pifont,multirow,threeparttable}

\usepackage{acronym}
\usepackage{mathrsfs}
\usepackage{gensymb}
\usepackage{hyperref}
\usepackage{lineno}  
\leftlinenumbers               

\begin{document} 


\title{Jet precession in gamma-ray bursts: The roles of fallback accretion and disk dynamics}

\author{Yun-Peng Li \inst{\ref{inst1}}
\and Da-Bin Lin \inst{\ref{inst1}}
\and Guo-Yu Li \inst{\ref{inst1}}
\and En-Wei Liang\inst{\ref{inst1}}}

\institute{Guangxi Key Laboratory for Relativistic Astrophysics, School of Physical Science and Technology, Guangxi University, Nanning 530004, China\label{inst1}
\\
\email{lindabin@gxu.edu.cn,lypzhw@st.gxu.edu.cn}
}

\date{Received XXX / Accepted XXX}

\abstract{
The precession phenomenon of the jet in a gamma-ray burst (GRB) is a key probe of the physics of the central engine.  Previous studies generally assumed a fixed precession period when analysing the temporal profiles in GRBs; however, the dynamic evolution of the fallback process and accretion disk can significantly affect the precession behaviour.
In this work we present a jet precession model that incorporates the co-evolution of fallback accretion and the central black hole (BH). Our model demonstrates that the jet precession period initially decreases rapidly during the early fallback phase and subsequently increases nearly linearly as the disk evolves. 
We find that a higher accretion disk viscosity and a slower BH spin lead to longer precession periods and faster precession period growth rates, and that the geometric structure of the precession system modulates the pulse amplitude of the light curve.
By fitting the model to observational data of GRBs with multi-pulse structures, we show that jet precession can naturally explain the increasing pulse intervals and broadened pulse widths observed in both long and short GRBs.}


\titlerunning{Jet precession in gamma-ray bursts}
\authorrunning{Yun-Peng Li et al.}

\maketitle

\section{Introduction}\label{sec:intro}

Gamma-ray bursts (GRBs) are among the most energetic events in the Universe. They are traditionally classified as long GRBs ($t_{\rm 90}\gtrsim 2\,{\rm s}$) or short GRBs ($t_{\rm 90}\lesssim 2\,{\rm s}$) based on the duration of their prompt emission (\citealp{1993ApJ...413L.101K}).
However, recent observations — such as the detection of a short GRB (GRB 200826A) associated with a supernova and two long GRBs (GRB 211211A and GRB 230307A) associated with kilonovae — have challenged this duration-based classification scheme (\citealp{2021NatAs...5..917A,2021NatAs...5..911Z,2022Natur.612..232Y,2022Natur.612..223R,2022Natur.612..228T,2022ApJ...933L..22Z,2023arXiv231007205Y,2024Natur.626..737L,2024Natur.626..742Y,2025NSRev..12E.401S}).
An alternative classification, based on a different physical origin, divides GRBs into Type~I (compact star merger GRBs) and Type~II (massive star GRBs; \citealp{2006Natur.444.1010Z,2009ApJ...703.1696Z,2025JHEAp..45..325Z}). 
Type~I GRBs are consistent with the merger of two compact objects, such as two neutron stars (NSs) or a NS and a black hole (BH), and are often accompanied by gravitational waves or kilonova emission (\citealp{1986ApJ...308L..43P,1989Natur.340..126E,1991AcA....41..257P,1992ApJ...395L..83N}). 
In contrast, Type II GRBs originate from the core collapse of massive stars (collapsars) and are typically associated with broad-lined Type Ic supernovae  (\citealp{1993ApJ...405..273W,1998ApJ...494L..45P,1999ApJ...524..262M}). 
In both scenarios, the central engine is thought to be a rotating BH or a millisecond magnetar surrounded by an accretion disk (\citealp{1989ApJ...346..847C,1992Natur.357..472U,1999ApJ...518..356P,2001ApJ...552L..35Z,2014MNRAS.438..240G,2018ApJ...857...95M,2020GReGr..52..108B,2024ApJ...968..104L,2025ApJ...985..195L}).
When the central engine is an accreting BH, a misalignment between the angular momentum of the fallback material and the BH spin can result in a tilted accretion disk. In such cases, the Lense–Thirring torque induces a precession of the disk, a phenomenon known as Lense–Thirring precession (\citealp{1918PhyZ...19..156L}). 
Consequently, the ultra-relativistic jet powered by the Blandford–Znajek (BZ) mechanism, the Blandford-Payne (BP) mechanism, or the neutrino-antineutrino annihilation mechanism launched from the central engine is also expected to precess (\citealp{1977MNRAS.179..433B,1982MNRAS.199..883B,1992MNRAS.257P..29M,2007A&A...468..563L,2010A&A...516A..16L,2012ChPhB..21f9801L}).

Jet precession has already been considered in the context of GRBs. Early studies (\citealp{2006A&A...454...11R,2007A&A...468..563L,2013PhRvD..87h4053S}) assumed that the post-merger accretion disk behaves as a thick disk that undergoes solid-body precession via the Lense-Thirring effect, which thus modulates the jet power. Regarding the thin disk scenario, \cite{2010A&A...516A..16L} proposed that the inner region of an inclined disk undergoes a Bardeen-Petterson warp, aligning its normal with the BH spin, and precesses around the total angular momentum vector. These jet precession models have been employed to explain various features observed in GRB light curves. For instance, \cite{2022ApJ...931L...2W} suggest that GRB~200826A is a Type~II GRB from the collapse of a massive star, with its $\sim 1$ s prompt emission serving as a precursor, but that the main emission was missed because jet precession shifted the radiation away from our line of sight. Similarly, \cite{2023ApJ...945...17G} explain that repeating temporal profiles in a GRB prompt light curves via a precessing jet, and \cite{2023ApJ...955...98L} attribute the delayed quasi-periodic oscillation behaviour in the Type~I GRB~130310A to jet precession in a BH--NS merger. In addition, \cite{2023RAA....23k5023Z} applied a jet precession and nutation model to GRB~220408B, identifying precession and nutation periods of $18.4 \pm 0.2$ s and $11.1 \pm 0.2$ s, respectively.

Previous studies have assumed a fixed precession period when analysing the temporal profiles of GRB light curves. However, \cite{2013PhRvD..87h4053S} found that the jet precession period can be influenced by the expansion of the outer boundary of the accretion disk, as described by the self-similarly evolving disk solution. In this study we investigated the evolutionary behaviour of jet precession periods by considering the co-evolution of fallback accretion and BHs. We provide temporal profiles of the corresponding light curves and compare them with observational candidates.

The paper is organized as follows. In Sect.~\ref{sec2} we present the jet precession model that incorporates the co-evolution of fallback accretion and BHs. Section~\ref{sec3} discusses the results of our model and their comparison with observations. Finally, conclusions and a discussion are given in Sect.~\ref{sec4}.

\section{Method} \label{sec2}   
\subsection{Co-evolution of the accretion disk and BH} \label{sub2:1}
After the core collapse of a massive star or the merger of a BH-NS, a newly born BH can remain, with the bound debris subsequently falling back to form an accretion disk. The fallback rate can be described as (\citealp{2018ApJ...857...95M})
\begin{equation}
        \dot{M}_{\rm fb}=\frac{2M_{\rm fb}}{3t_{\rm fb}(1+t/t_{\rm fb})^{5/3}},
\end{equation}
where $M_{\rm fb}$ and $t_{\rm fb}$ are the total fallback mass and characteristic fallback timescale, respectively. The fallback material contributes to the radial surface density $\Sigma(r,t)$ of the accretion disk, whose evolution is governed by (\citealp{2008bhad.book.....K})
\begin{equation}
        \frac{\partial \Sigma}{\partial t} = \frac{1}{2\pi r} \frac{\partial \dot{M}}{\partial r} + \dot{\Sigma}_{\rm fb},
\end{equation}
where $\dot{\Sigma}_{\rm fb}$ is the rate of surface density increase due to fallback material. The material is assumed to fall onto the disk at around the radius $r_{\rm fb}$. The spatial distribution of $\dot{\Sigma}_{\rm fb}$ is modelled as a Gaussian profile:   
\begin{equation}
        \dot{\Sigma}_{\rm fb}=\frac{(1-f_{\rm w})\dot{M}_{\rm fb}}{2\pi rr_{\rm fb}}A\exp\left[-(\frac{r-r_{\rm fb}}{r_{\rm fb}/4})^2\right],
\end{equation}
where $f_{\rm w}$ is the mass fraction lost via disk wind (\citealp{2023ApJ...957L...9T}), and $A$ is the dimensionless normalization factor determined by $\int_{r_{\rm in}}^{r_{\rm out}} \dot{\Sigma}_{\rm fb} \cdot 2\pi r \, dr = (1 - f_{\rm w}) \dot{M}_{\rm fb}$, with $r_{\rm in}$ and $r_{\rm out}$ being the disk's inner and outer radii.  

Furthermore, the accretion rate, $\dot{M}$, can be estimated as (e.g. Eq. $3.15$ in \citealt{2008bhad.book.....K})
\begin{equation}\label{Eq4}
        \dot{M}(r,t)=6\pi\sqrt{r}\frac{\partial }{\partial r}(\nu\Sigma r^{1/2}).       
\end{equation}   
Here, $\nu=\alpha c_{\rm{s}}H$, where $\alpha$ is the viscosity parameters, $c_{\rm{s}}$ is the sound speed of gas, and $H$ is the half thickness of the disk at radius $r$.  
The values of $c_{\rm{s}}$ and $H$ are determined by the advection factor $f_{\rm{adv}}(=Q_{\rm{adv}}^-/Q_{\rm{adv}}^+)$ of the accretion flow, i.e. $c_{\rm{s}}\sim v_{\rm{\phi}}\sqrt{f_{\rm{adv}}}$ and $H/r\approx \sqrt{f_{\rm{adv}}}$, where $Q_{\rm{adv}}^-$ and $Q_{\rm{adv}}^+$ are the factors of the advection cooling and viscous heating in the accretion flow, and $v_{\rm{\phi}}$ is the Kepler’s rotation velocity of gas around BH. 
For the sake of simplicity in our analysis, we assumed that $H/r$ remains constant (\citealp{2013PhRvD..87h4053S}).

Jet formation is intimately linked with the accretion process.
A common mechanism invoked to power BH jets is via BZ process (\citealp{1977MNRAS.179..433B}).
The BZ power is related to the mass and spin of the central BH (\citealp{2000ApJ...534L.197L,2013ApJ...765..125L,2013ApJ...767L..36W}), which can be given by
\begin{equation}
        L_{\rm BZ}=1.7\times10^{50} {\rm erg\,s^{-1}} a_{\rm BH}(\frac{M_{\rm BH}}{M_{\odot}})^2B_{\rm BH,15}^2F(a_{\rm BH}),
\end{equation}
where $a_{\rm BH} \in [0,1]$ is the dimensionless BH spin parameter, $M_{\rm BH}$ is the BH mass, $B_{\rm BH,15}=B_{\rm BH}/10^{15}\, \rm G$ is the magnetic field strength threading the BH horizon, and $F(a_{\rm BH})=[(1+q^2)/q^2][(q+1/q) {\rm arctan}q-1]$ with $q=a_{\rm BH}/(1+\sqrt{1-a_{\rm BH}^2})$. Through the balance between the magnetic pressure on the horizon and the ram pressure of the innermost part of the accretion flow, the magnetic field can be estimated by (\citealp{2011MNRAS.418L..79T,2017ApJ...849...47L})
 \begin{equation}
        \frac{B_{\rm BH}^2}{8\pi}=P_{\rm ram}\sim \rho c^2 \sim \frac{\dot{M}(r_{\rm in})c}{4\pi r_{\rm H}^2},
 \end{equation}
where $r_{\rm H}=(1+\sqrt{1-a_{\rm BH}^2})r_{\rm g}$ is the radius of the BH horizon, $r_{\rm g}=GM_{\rm BH}/c^2$ is the gravitational radius, $G$ and $c$ denote the gravitational constant and the speed of light, respectively. Additionally, $\dot{M}(r_{\rm in})$ can be given by Eq.~(\ref{Eq4}). Therefore, the BZ power can be given as (\citealp{2022ApJ...936L..10Z})
 \begin{equation}
        L_{\rm BZ}=9.3\times10^{53} {\rm erg\,s^{-1}} \frac{a_{\rm BH}^2F(a_{\rm BH})}{(1+\sqrt{1-a_{\rm BH}^2})^2} \frac{\dot{M}(r_{\rm in})}{M_{\odot}\,{\rm s}^{-1}}.
 \end{equation}
 
Then, the evolution process for mass and spin of a spinning BH can be written as (\citealp{2014ApJ...781L..19H,2022ApJ...936L..10Z})
\begin{equation} 
        \frac{dM_{\rm BH}}{dt}=\dot{M}(r_{\rm in})e_{\rm ms}-L_{\rm BZ},
\end{equation}
\begin{equation} \label{Eq9}
        \frac{dJ_{\rm BH}}{dt}=\dot{M}(r_{\rm in})l_{\rm ms}-2L_{\rm BZ}/\Omega_{\rm H},
\end{equation}
where $J_{\rm BH}=GM_{\rm BH}^2a_{\rm BH}/c$ is the angular momentum of BH, $\Omega_{\rm H}=ca_{\rm BH}/(2r_{\rm H})$ is the angular velocity of the BH horizon, $e_{\rm ms}=1/\sqrt{3\hat{r}_{\rm ms}}(4-3a_{\rm BH}/\sqrt{\hat{r}_{\rm ms}})$ and $l_{\rm ms}=2\sqrt{3}GM_{\rm BH}(1-2a_{\rm BH}/(3\sqrt{\hat{r}_{\rm ms}}))/c$ are the specific energy and angular momentum at the marginally stable orbit, with $\hat{r}_{\rm ms}=3+Z_2-[(3-Z_1)(3+Z_1+2Z_2)]^{1/2}$ (\citealp{2012ApJ...760...63L,2013ApJ...767L..36W}).
In a precessing system, a misalignment between the angular momentum of the accreted material and that of the BH can reorient the BH spin, and thus Eq.~(\ref{Eq9}) should be treated as a vector. However, for systems with $M_{\rm fb} \sim 10^{-1}\,M_{\odot}$, the component of the accreted angular momentum perpendicular to the BH spin has a negligible effect on the spin orientation\footnote{For a representative case with $M_{\rm BH}=5\,M_\odot$, $a_{\rm BH}=0.8$, and $M_{\rm fb}=0.1\,M_\odot$, the ratio of the total angular momentum of the accreted material ($J_{\rm acc,tot}$) to that of the BH ($J_{\rm BH}$) is $\sim10^{-2}$. When the angle between the two angular momentum vectors is small ($<10^\circ$), the component of $J_{\rm acc,tot}$ perpendicular to the BH spin is only $\sim 10^{-3}\,J_{\rm BH}$, implying a deflection in the spin orientation of $\arctan(10^{-3}) \simeq 0.06^\circ$ over the entire accretion episode. It is worth noting that if the mass of the accreted material becomes comparable to $M_{\rm BH}$, a full vectorial treatment combining $J_{\rm BH}$ and the angular momentum of the inflowing matter becomes necessary.}. Therefore, we simply treated the evolution of angular momentum for accretion system as a scalar.

\subsection{Jet precession} \label{sub2:2}

During a BH–NS merger, the orbital plane of the binary can be misaligned with the BH's spin axis, resulting in the formation of an accretion disk that is initially tilted with respect to the BH's spin (\citealp{2011PhRvD..83b4005F}). In addition, anisotropic explosions of massive stars can also lead to a misalignment between the angular momentum of the resulting disk and that of the BH (\citealp{2010A&A...516A..16L,2022ApJ...931L...2W}). Similarly, core collapse induced by misaligned massive star mergers can also yield a tilted accretion disk. 
Such tilted disks are subject to Lense–Thirring precession around the BH spin axis (\citealp{2007ApJ...668..417F,2012PhRvL.108f1302S}), with the associated relativistic jets precessing coherently with the disk (\citealp{2018MNRAS.474L..81L}). The inner disk region aligns with the BH equatorial plane through the Bardeen–Petterson effect (\citealp{1975ApJ...195L..65B}) for thin disks, while thick disks ($H/r \gtrsim \alpha$) exhibit near-rigid-body precession, as demonstrated by analytic studies~(\citealp{1983MNRAS.202.1181P,1995ApJ...438..841P}), hydrodynamic simulations~(\citealp{1997MNRAS.285..288L,2000MNRAS.315..570N}), and general relativistic magnetohydrodynamic simulations~(\citealp{2007ApJ...668..417F,2011ApJ...730...36D,2018MNRAS.474L..81L}). In the following, we model the precession period $T_{\rm prec}$ under the assumption of the thick accretion disk.

The Newtonian rigid-body precession period is 
\begin{equation}
   T_{\rm prec}=2\pi\,{\rm sin}\psi(J_{\rm d}/\mathcal{N}_{\rm d}),
\end{equation} 
the total angular momentum of the disk (\citealp{2013PhRvD..87h4053S}) can be given as 
\begin{equation}
        J_{\rm d}=2\pi\int_{r_{\rm in}}^{r_{\rm out}}{\Sigma(r,t) \Omega(r) r^3 dr},
\end{equation}
and the radially integrated Lense–Thirring torque can be described as
\begin{equation}
        \mathcal{N}_{\rm d}=4\pi \frac{G^2 M_{\rm BH}^2 a_{\rm BH}}{c^3}\int_{r_{\rm in}}^{r_{\rm out}}{\Sigma(r,t) \Omega(r) dr}.
\end{equation}
The orbital frequency $\Omega(r) = \sqrt{GM_{\rm BH}/r^3}$ follows the Keplerian prescription (\citealp{2013PhRvD..87h4053S}).  
The inner radius of the disk $r_{\rm in}$ corresponds to the innermost stable circular orbit (ISCO) for aligned disks and to the innermost stable spherical orbit (ISSO) for misaligned cases (\citealp{2013MNRAS.435.1809S}). Given the nearly negligible difference between ISCO and ISSO at precession angles $\theta_{\rm p} < 30^\circ$ (\citealp{2023ApJ...957L...9T}), we adopted $r_{\rm in}=r_{\rm ISCO}$ as defined by \cite{1972ApJ...178..347B}:
\begin{equation}
        r_{\rm ISCO}=(3+Z_2-[(3-Z_1)(3+Z_1+2Z_2)]^{1/2})r_{\rm g},
\end{equation}
where $Z_1=1+(1-a_{\rm BH}^2)^{1/3}[(1+a_{\rm BH})^{1/3}+(1-a_{\rm BH})^{1/3}]$, and $Z_2=(3a_{\rm BH}^2+Z_1^2)^{1/2}$.

\begin{figure}[htbp]
        \centering
        \includegraphics[width=0.5\textwidth]{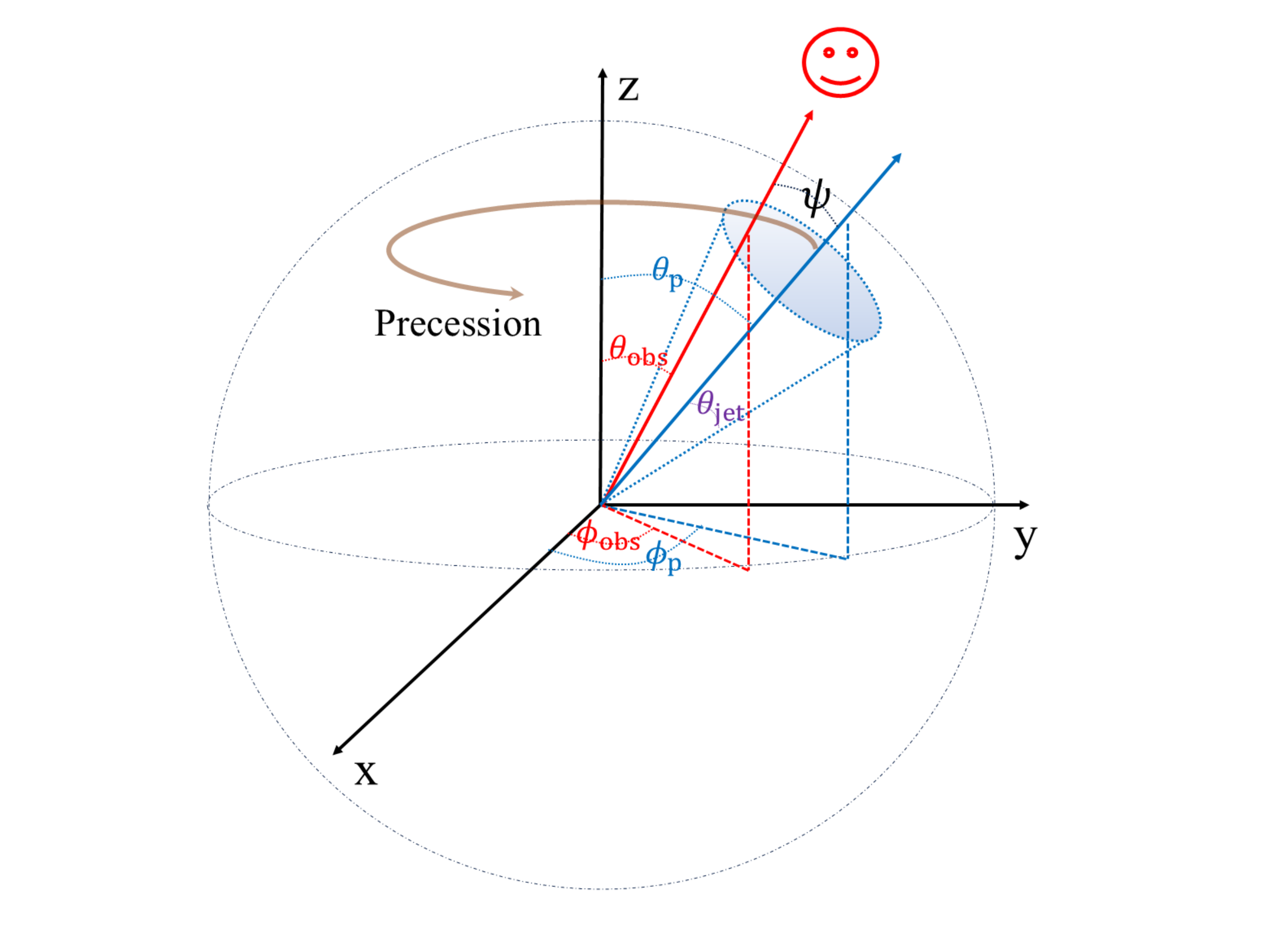}
        
\caption{Schematic illustration of a BH-driven precessing jet. The blue arrows represent the jet orientation, which precesses around the $Z$-axis with a precession angle $\theta_{\rm p}$ and an azimuthal angle $\phi_{\rm p}$. The red arrow represents the observer's line of sight, defined by an inclination angle $\theta_{\rm obs}$ and an azimuthal angle $\phi_{\rm obs}$. The jet opening angle is denoted by $\theta_{\rm jet}$, while the brown arrow indicates the direction of precession.}

        \label{fig1: Schematic illustration}
\end{figure}

As illustrated in Fig.~\ref{fig1: Schematic illustration}, the jet precesses around the $\mathrm{Z}$-axis with a precession angle $\theta_{\rm p}$. The angle between the observer’s line of sight and the $\mathrm{Z}$-axis is $\theta_{\rm obs}$, while the azimuthal angle around the $\mathrm{Z}$-axis is $\phi_{\rm obs}$. The jet opening angle is denoted by $\theta_{\rm jet}$. The instantaneous angle between the observer’s line of sight and the jet axis is
\begin{equation}
        \psi(t)=\cos^{-1}\Big[\cos\theta_{\rm p}\cos\theta_{\rm obs}+\sin\theta_{\rm p}\sin\theta_{\rm obs}\cos\big(\phi_{\rm p}(t)-\phi_{\rm obs}\big)\Big],
\end{equation}
where $\phi_{\rm p}(t)$, the azimuthal angle of the jet, evolves according to $d\phi_{\rm p}/ dt=2\pi/T_{\rm prec}$.

To model the light curves, we adopted a Gaussian-structured jet model (\citealp{2002ApJ...571..876Z}), which is consistent with observations of GRB~170817A (\citealp{2018MNRAS.478..733L,2018PhRvL.120x1103L,2018Natur.561..355M,2018MNRAS.478L..18T,2018ApJ...863...58X}). 
For a structured jet, the observed light curve should be calculated by integrating the emission from different jet regions with appropriate Doppler shift corrections (\citealp{2025arXiv250501606C}). However, when the viewing angle is not much larger than the jet's characteristic Gaussian angle (i.e. $\theta_{\rm p} - \theta_{\rm obs} \lesssim \theta_{\rm jet}$; see Fig.~\ref{fig1: Schematic illustration}), the total observed emission does not differ significantly from the simplified scenario where only the line-of-sight contribution is considered. In this work, we focus solely on on-beam configurations (i.e. $\theta_{\rm p} - \theta_{\rm obs} \lesssim \theta_{\rm jet}$), and the time-dependent luminosity can be simply expressed as (\citealp{2023ApJ...945...17G})  
\begin{equation}
        L(t) = L_0(t) f(\theta = \psi(t)),
\end{equation}
where $L_0 =  \eta \dot{M}(r_{\rm in}) c^2$, with $\eta = 0.1$ representing accretion energy conversion efficiency (\citealp{2004MNRAS.351..169M,2004MNRAS.354.1020S}). The angle-dependent luminosity profile can be written as
\begin{equation}
        f(\theta) = e^{-\theta^2/(2\theta_{\rm c}^2)},
\end{equation}
with $\theta_{\rm c}$ the jet core angle.
For Type~II GRBs, the relativistic jet initially interacts with the progenitor's stellar envelope, carving out a funnel and potentially producing a reactive force that can suppress jet precession. However, if the jet power is sufficiently strong and the envelope radius is relatively small, the precession signature can still be preserved. In this work we assumed that the jet retains its precession characteristics after breaking out of the envelope.

\section{Results} \label{sec3}

\subsection{Evolution of the precession period in jet precession} \label{sub3:1}

\subsubsection{Effects of accretion disk viscosity and thickness\,} 
\begin{figure*}[htbp]
        \centering
        \includegraphics[width=0.49\textwidth]{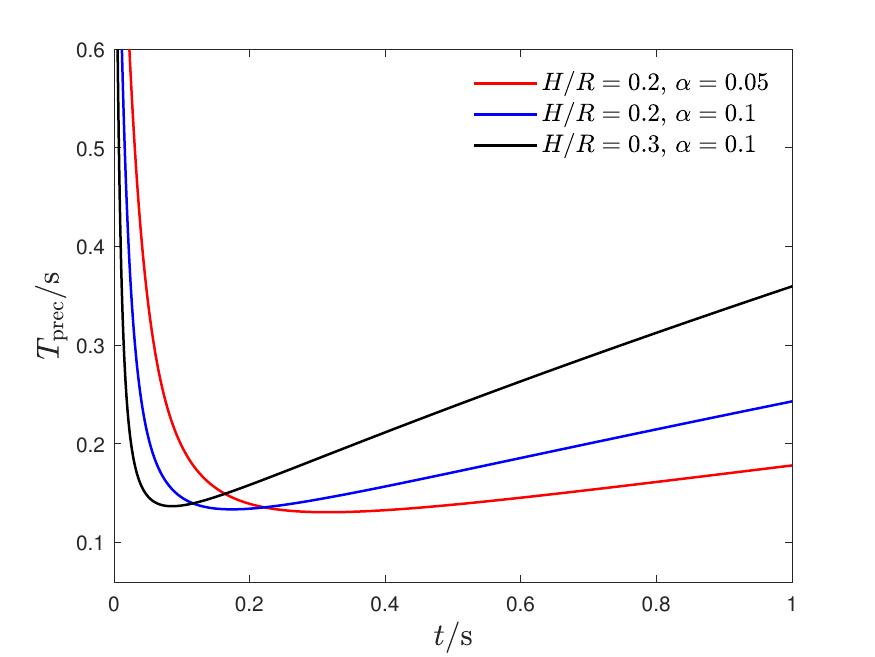}
        \includegraphics[width=0.49\textwidth]{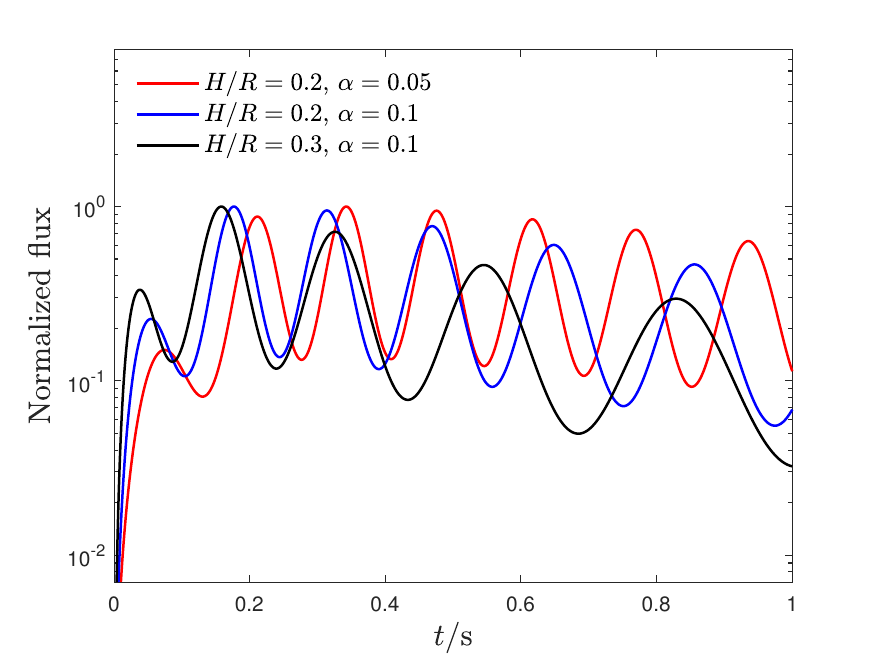}
        \caption{Evolution of the precession period ($T_{\rm prec}$; left panel) and the normalized flux (right panel) with time. The blue, red, and black curves correspond to different accretion disk parameters: $H/R=0.3$, $\alpha=0.05$; $H/R=0.2$, $\alpha=0.1$; and $H/R=0.3$, $\alpha=0.1$, respectively. The initial parameters ($M_{\rm BH,0}, a_{\rm BH,0}, M_{\rm fb}, R_{\rm fb}, t_{\rm fb}, \theta_{\rm p}, \theta_{\rm c}, \theta_{\rm obs}, \phi_{\rm p}, \phi_{\rm obs,0}$) are set as ($5M_{\odot}, 0.8, 0.1M_{\odot}, 30r_{\rm g}, 0.1{\rm s}, 5^{\circ}, 5^{\circ}, 5^{\circ}, 0^{\circ}, 0^{\circ}$), respectively.}
        
    \label{fig2: HR}
\end{figure*}

We first investigated the evolution of the precession period under varying viscosity parameters, $\alpha,$ and disk half-thickness ratios, $H/R$. With initial parameters ($M_{\rm BH,0}, a_{\rm BH,0}, M_{\rm fb}, R_{\rm fb}, t_{\rm fb}, \theta_{\rm p}, \theta_{\rm c}, \theta_{\rm obs}, \phi_{\rm p}, \phi_{\rm obs,0}$) = ($5M_{\odot}, 0.8, 0.1M_{\odot}, 30r_{\rm g}, 0.1\,{\rm s}, 5^{\circ}, 5^{\circ}, 5^{\circ}, 0^{\circ}, 0^{\circ}$), Fig.~\ref{fig2: HR} shows the corresponding light curves and the evolution trend of $T_{\rm prec}$. 
The left panel of Fig.~\ref{fig2: HR} reveals a characteristic two-phase evolution for $T_{\rm prec}$: an initial rapid decline in precession period during initial fallback of the materials, followed by a nearly linear increase ($T_{\rm prec} \propto t$). This long-term growth for $T_{\rm prec}$ is dominated by the outward expansion of the disk's surface density. Both the disk thickness and the viscosity parameter significantly modulate the evolution slope of $T_{\rm prec}$. Thicker disks (larger $H/R$) and larger viscosity parameters ($\alpha$) lead to a steeper increase in $T_{\rm prec}$. The right panel demonstrates that a more rapid increase in $T_{\rm prec}$ corresponds to higher pulse frequencies in light curves over fixed time intervals. Additionally, the time interval between adjacent pulses increases gradually with time, and the pulse width also broadens progressively.

\subsubsection{Effects of dynamical parameters and the angular parameter\,}
\begin{figure*}[htbp]
        \centering
        \includegraphics[width=0.49\textwidth]{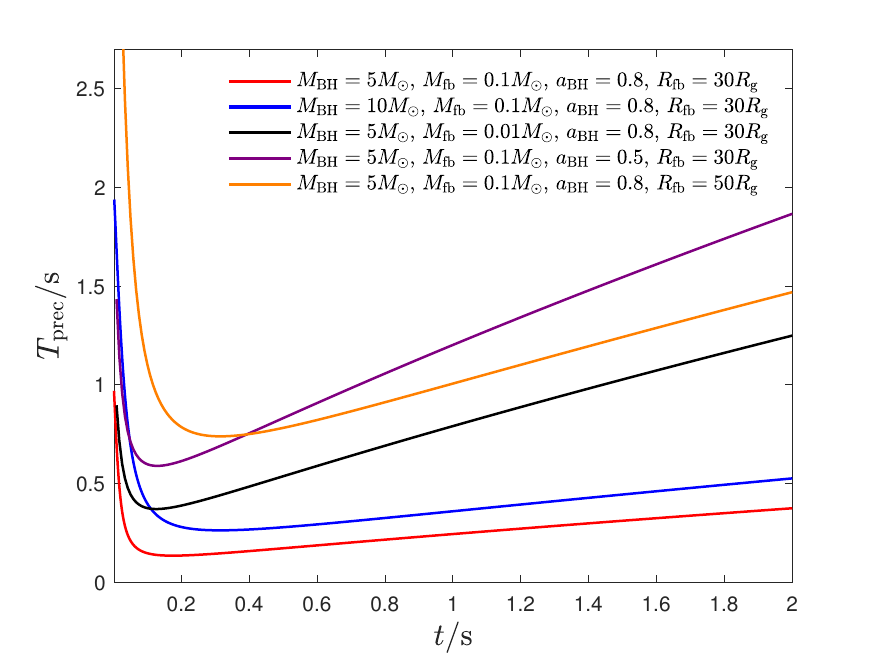}
        \includegraphics[width=0.49\textwidth]{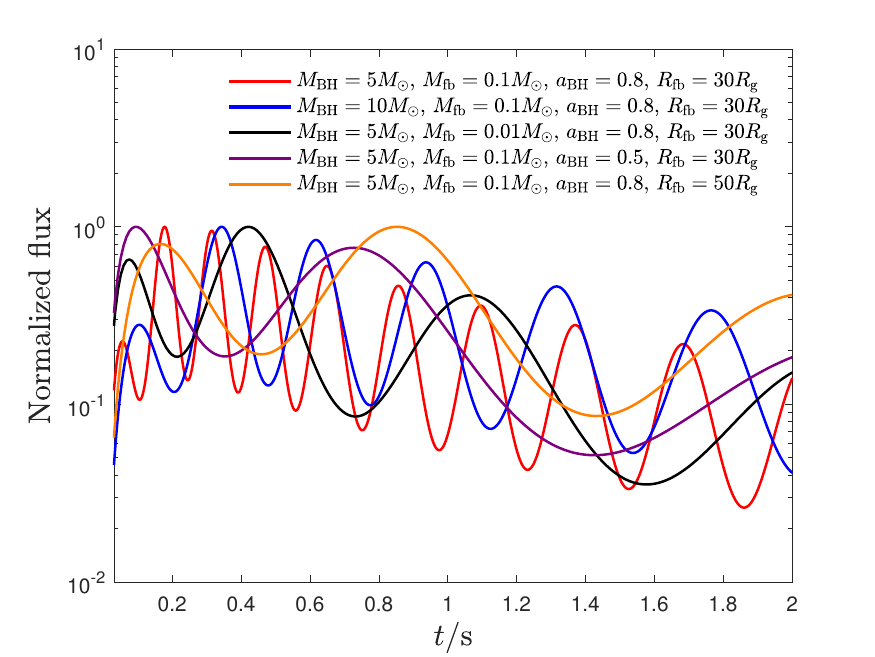}
        \caption{Evolution of the precession period ($T_{\rm prec}$; left panel) and the normalized flux (right panel) with time. Different colours represent different parameter combinations. $H/R=0.2$, $\alpha=0.1$, and $t_{\rm fb}=0.1\,\rm s$ are fixed, and the initial parameters ($\theta_{\rm p}, \theta_{\rm c}, \theta_{\rm obs}, \phi_{\rm p}, \phi_{\rm obs,0}$) are set as ($5^{\circ}, 5^{\circ}, 5^{\circ}, 0^{\circ}, 0^{\circ}$).}
        
        \label{fig3: different parameters}
\end{figure*}

\begin{figure}[htbp]
        \centering
        \includegraphics[width=0.5\textwidth]{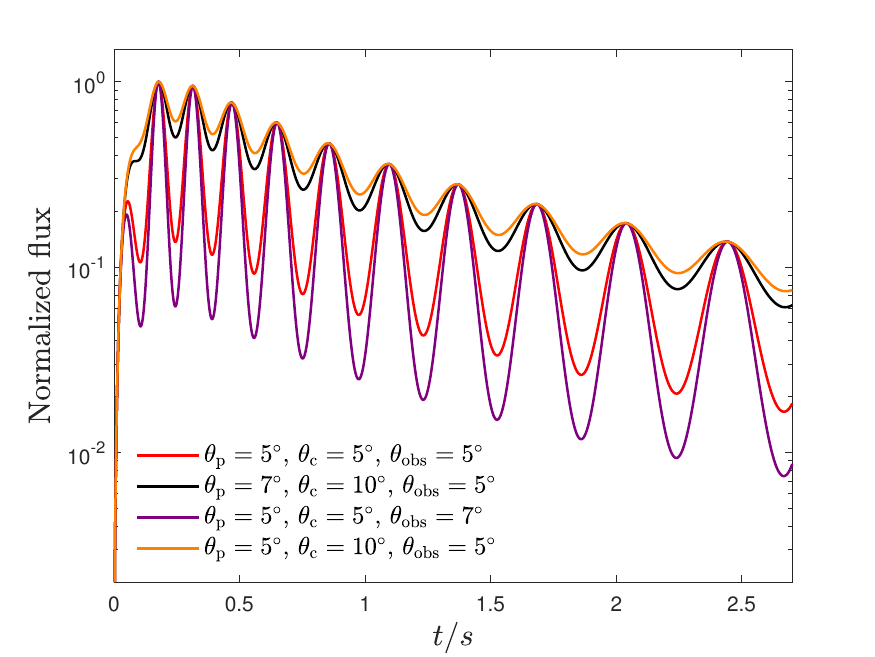}
        
\caption{Normalized flux under different angle parameter combinations. Different colours represent different angle parameter combinations. $H/R=0.2$, $\alpha=0.1$, and $t_{\rm fb}=0.1\,\rm s$ are fixed, and the initial parameters ($M_{\rm BH,0}, a_{\rm BH,0}, M_{\rm fb}, R_{\rm fb}$) are set as ($5M_{\odot}, 0.8, 0.1M_{\odot}, 30r_{\rm g}$).}
        \label{fig4: angle}
\end{figure}

We next analysed how dynamical parameters influence the evolution of the precession period. By fixing $ H/R=0.2 $, $ \alpha=0.1 $, and $ t_{\rm fb}=0.1\,\mathrm{s} $, and setting initial angular parameters $( \theta_{\rm p}, \theta_{\rm c}, \theta_{\rm obs}, \phi_{\rm p}, \phi_{\rm obs,0} ) = ( 5^{\circ}, 5^{\circ}, 5^{\circ}, 0^{\circ}, 0^{\circ} )$, Fig.~\ref{fig3: different parameters} presents a comparison of the precession period $ T_{\rm prec} $ and light curves for different dynamical parameters.
The left panel shows that systems with lower spins ($a_{\rm BH,0}$), smaller fallback masses ($M_{\rm fb}$), and more distant fallback radii ($R_{\rm fb}$) exhibit faster $T_{\rm prec}$ growth rates. Additionally, higher $M_{\rm BH,0}$, lower $a_{\rm BH,0}$, higher fallback masses ($M_{\rm fb}$), and more distant fallback radii ($R_{\rm fb}$) generally produce longer precession periods. Corresponding light curves in the right panel illustrate how these dynamical differences translate to observational characteristics.
Finally, we examined angular parameter impacts using fixed $H/R=0.2$, $\alpha=0.1$, $t_{\rm fb}=0.1\,\rm s$, and initial dynamical parameters ($M_{\rm BH,0}, a_{\rm BH,0}, M_{\rm fb}, R_{\rm fb}$) = ($5M_{\odot}, 0.8, 0.1M_{\odot}, 30r_{\rm g}$). Figure~\ref{fig4: angle} demonstrates that angular combinations primarily govern pulse amplitudes. Specific orientations between precession angle ($\theta_{\rm p}$), jet core angle ($\theta_{\rm c}$), and observer angle ($\theta_{\rm obs}$) play a critical role in determining the pulse amplitudes observed in the light curves.

\subsection{Case study} \label{sub3:2}
Our jet precession model demonstrates that parameter combinations control both pulse frequencies and amplitudes in GRB light curves. Importantly, the temporal evolution of the precession period produces progressively wider pulse intervals and broader pulse widths, a key observational signature. This prediction aligns with the observed rarity of quasi-periodic signals in Type~I GRBs (\citealp{2023ApJ...955...98L}), as an increasing $T_{\rm prec}$ would smear out periodic features over typical burst durations.
Notably, jet precession has also been proposed as a central mechanism in Type~II GRBs (\citealp{1996ApJ...473L..79B,1999ApJ...520..666P,2006A&A...454...11R,2010A&A...524A...4R}). In the following, we analyse Type~I and Type~II GRBs that exhibit multi-pulse behaviour and interpret them within the framework of our precession model by fitting their temporal profiles.

\subsubsection{Fitting for GRB~171126A\,}  
\begin{figure}[htb]
        \centering
    \includegraphics[width=0.5\textwidth]{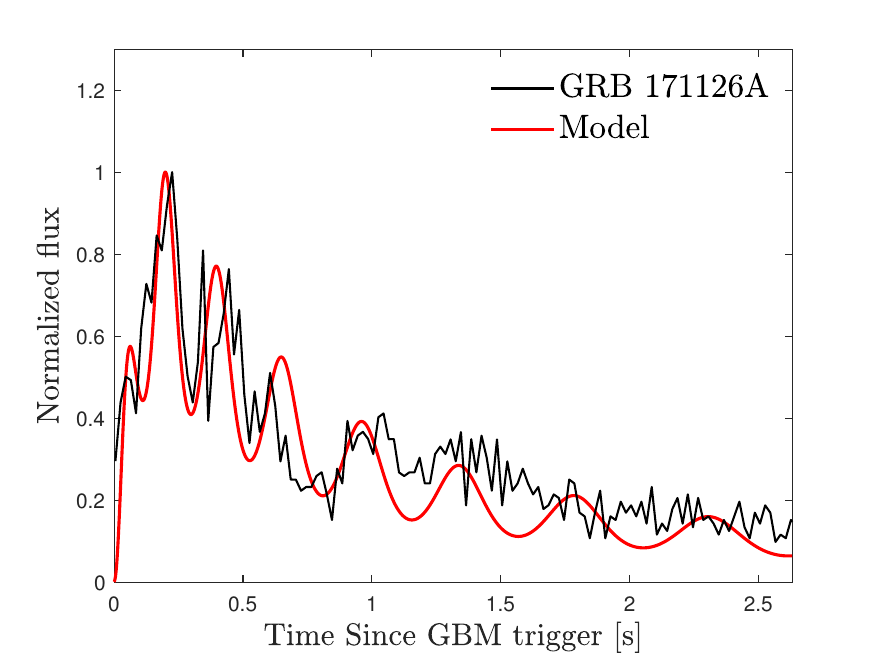}
        
        \caption{Normalized light curve of GRB~171126A and the jet precession model. The accretion disk parameters ($H/R,\, \alpha$)=($0.3,\, 0.05$), the dynamic parameter ($M_{\rm BH,0}, a_{\rm BH,0}, M_{\rm fb}, R_{\rm fb}, t_{\rm fb}$) = ($4M_{\odot}, 0.65, 0.1M_{\odot}, 26r_{\rm g}, 0.1\rm s$), and angle parameters ($\theta_{\rm p}, \theta_{\rm c}, \theta_{\rm obs}, \phi_{\rm p,0}, \phi_{\rm obs}$) = ($5^{\circ}, 8^{\circ}, 5^{\circ}, 0^{\circ}, 40^{\circ}$) are adopted to fit the observation.}
        \label{fig5: short GRB}

\end{figure}

In our survey of Type~I GRBs with multi-pulse characteristics, GRB~171126A is found to exhibit increasing pulse widths and intervals over time. We applied the jet precession model to fit its light curve (see Fig.~\ref{fig5: short GRB}).
The accretion disk parameters ($H/R,\, \alpha$)=($0.3,\, 0.05$), the dynamic parameter ($M_{\rm BH,0}, a_{\rm BH,0}, M_{\rm fb}, R_{\rm fb}, t_{\rm fb}$) = ($4M_{\odot}, 0.65, 0.1M_{\odot}, 26r_{\rm g}, 0.1\rm s$), and angle parameters ($\theta_{\rm p}, \theta_{\rm c}, \theta_{\rm obs}, \phi_{\rm p,0}, \phi_{\rm obs}$) = ($5^{\circ}, 8^{\circ}, 5^{\circ}, 0^{\circ}, 40^{\circ}$) were adopted to fit the observation. The model light curve reproduces the observed features to a certain degree, suggesting that jet precession may be responsible for the multi-pulse characteristics of GRB~171126A, and implying that this event likely originates from a BH--NS merger.

\subsubsection{Fitting for GRB~211129A\,}  
\begin{figure}[htbp]
        \centering
        \includegraphics[width=0.5\textwidth]{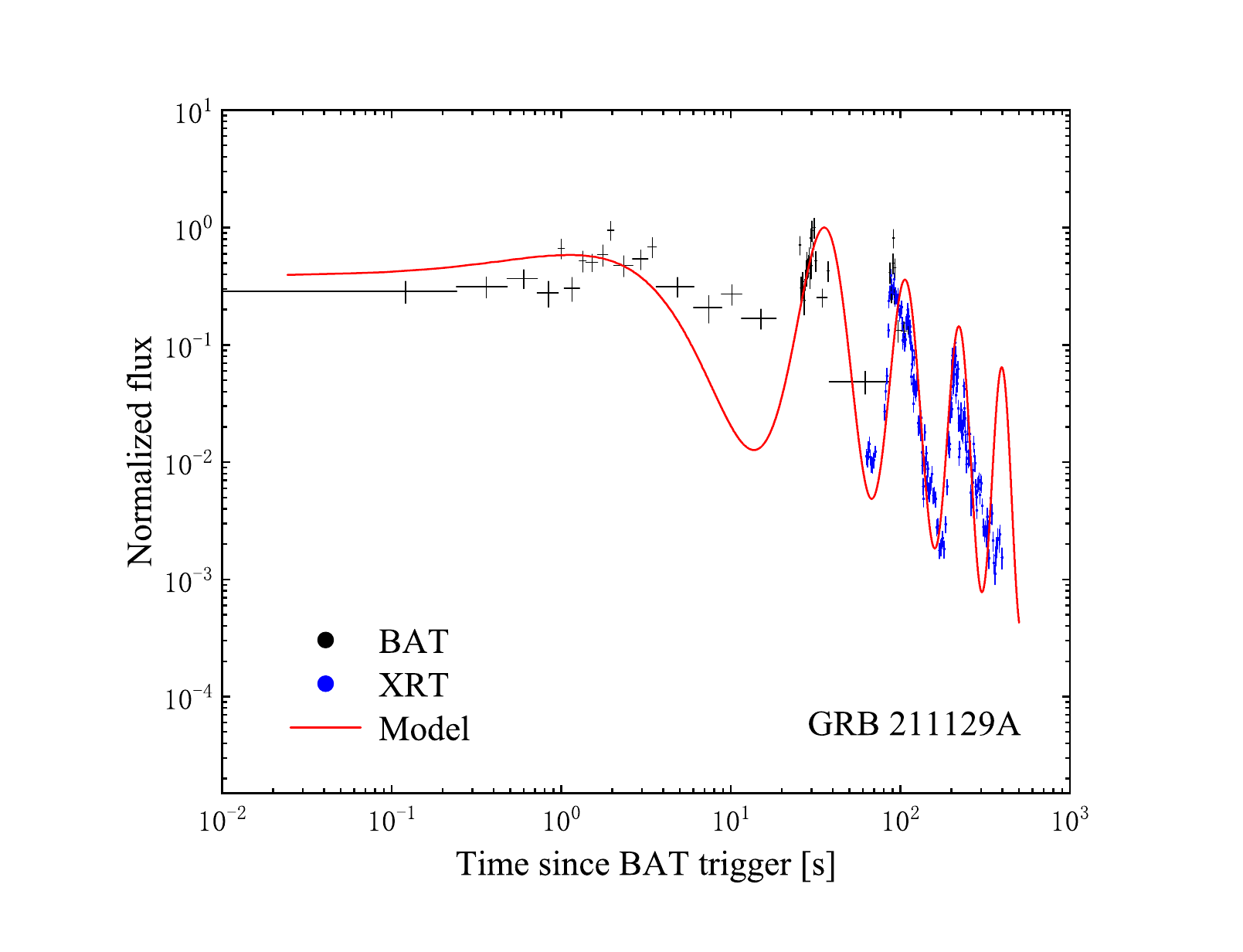}
        \caption{Normalized light curve of GRB~211129A and the jet precession model. The accretion disk parameters ($H/R,\, \alpha$)=($0.3,\, 0.1$), the dynamic parameter ($M_{\rm BH,0}, a_{\rm BH,0}, M_{\rm fb}, R_{\rm fb}, t_{\rm fb}$) = ($16M_{\odot}, 0.15, 0.1M_{\odot}, 50r_{\rm g}, 40\rm s$), and angle parameters ($\theta_{\rm p}, \theta_{\rm c}, \theta_{\rm obs}, \phi_{\rm p}, \phi_{\rm obs,0}$) = ($10^{\circ}, 5^{\circ}, 6^{\circ}, 0^{\circ}, 30^{\circ}$) are adopted to fit the light curve of GRB~211129A. }
        
        \label{fig6: long GRB}
\end{figure}

In the case of massive star collapse, anisotropic explosions of massive stars or core collapse induced by misaligned massive star mergers can lead to the formation of a jet precession system, resulting in multi-pulse light curves for Type~II GRBs. Recently, GRB~211129A, detected by \textit{Swift}, exhibited multiple pulses with increasing intervals and pulse widths over time. We propose that its multi-pulse features arise from jet precession. 
The fitting is performed using the following parameters: accretion disk parameters ($H/R,\, \alpha$)=($0.3,\, 0.1$), the dynamic parameter ($M_{\rm BH,0}, a_{\rm BH,0}, M_{\rm fb}, R_{\rm fb}, t_{\rm fb}$) = ($16M_{\odot}, 0.15, 0.2M_{\odot}, 50r_{\rm g}, 40\rm s$), and angle parameters ($\theta_{\rm p}, \theta_{\rm c}, \theta_{\rm obs}, \phi_{\rm p}, \phi_{\rm obs,0}$) = ($10^{\circ}, 5^{\circ}, 6^{\circ}, 0^{\circ}, 30^{\circ}$). The fitting result are shown in Fig.~\ref{fig6: long GRB}. 
The fitting results indicate that a low dimensionless spin $a_{\rm BH}$ is required. This suggests that the BZ mechanism might be inefficient in this case, and that additional jet generation mechanisms — such as neutrino annihilation (\citealp{2010A&A...516A..16L,2012ApJ...752...31S}) or the BP mechanism (\citealp{1982MNRAS.199..883B}) — may be needed to sustain the jet power.

\section{Discussion and conclusion} \label{sec4}

In this work we have developed a jet precession model that captures the co-evolution of the accretion disk and the BH, enabling us to study both the evolution of the jet precession period and the corresponding light curve characteristics. Our analysis shows that the precession period initially decreases rapidly due to the fallback of bound material. Once a stable accretion disk is established, however, the period increases linearly with time as a result of the outward diffusion of the disk's surface matter. Consequently, both the pulse interval and the pulse width of the light curve gradually increase with time.
Furthermore, the absolute value and evolution of the precession period, $T_{\rm prec}$, depend on several parameters: viscosity parameters ($\alpha$), disk half-thickness ratios ($H/R$), the BH mass ($M_{\rm BH,0}$), the BH spin ($a_{\rm BH,0}$), the fallback mass ($M_{\rm fb}$), and the fallback radius ($R_{\rm fb}$). In general, longer precession periods are associated with larger values of $\alpha$, $H/R$, and $M_{\rm BH,0}$; a lower initial BH spin ($a_{\rm BH,0}$); a lower fallback mass ($M_{\rm fb}$); and a larger fallback radius ($R_{\rm fb}$). In addition, the geometrical configuration — defined by the precession angle ($\theta_{\rm p}$), the jet core angle ($\theta_{\rm c}$), and the observer inclination ($\theta_{\rm obs}$) — determines the amplitude and modulation depth of the pulses.

Our model also provides a good match to the observational data of GRB~171126A and GRB~211129A, both of which exhibit multi-pulse characteristics with gradually increasing pulse intervals and widths. For the Type~I GRB GRB~171126A, we adopted the accretion disk parameters $H/R=0.3$ and $\alpha=0.05$, dynamic parameters $M_{\rm BH,0}=4\,M_{\odot}$, $a_{\rm BH,0}=0.65$, $M_{\rm fb}=0.1\,M_{\odot}$, $R_{\rm fb}=26\,r_{\rm g}$, and $t_{\rm fb}=0.1\,\rm s$, and angle parameters $\theta_{\rm p}=5^{\circ}$, $\theta_{\rm c}=8^{\circ}$, $\theta_{\rm obs}=5^{\circ}$, $\phi_{\rm p,0}=0^{\circ}$, and $\phi_{\rm obs}=40^{\circ}$. The well-matched light curves suggest that GRB~171126A originates from a BH--NS merger. For the Type~II GRB GRB~211129A, we adopted the accretion disk parameters $H/R=0.3$ and $\alpha=0.1$, dynamic parameters $M_{\rm BH,0}=16\,M_{\odot}$, $a_{\rm BH,0}=0.15$, $M_{\rm fb}=0.2\,M_{\odot}$, $R_{\rm fb}=50\,r_{\rm g}$, and $t_{\rm fb}=40\,\rm s$, and angle parameters $\theta_{\rm p}=10^{\circ}$, $\theta_{\rm c}=5^{\circ}$, $\theta_{\rm obs}=6^{\circ}$, $\phi_{\rm p}=0^{\circ}$, and $\phi_{\rm obs,0}=30^{\circ}$. The relatively low value of $a_{\rm BH}$ indicates that an additional jet energy source, such as neutrino annihilation or the BP mechanism, may be required to supplement the inefficient BZ jet power (\citealp{1982MNRAS.199..883B,2010A&A...516A..16L}). This finding is consistent with previous studies showing that BHs formed by the collapse of massive stars tend to have low spins (\citealp{2019ApJ...881L...1F}).

Finally, we note that several factors can mask the precession-induced pulse characteristics in observations. A long precession period can cause the pulse width to exceed the duration of the prompt emission, while a small precession angle can reduce the pulse amplitude, making the pulses less distinguishable. Moreover, the intrinsic variability of the jet may further obscure these features.

\begin{acknowledgements}
We thank the anonymous referee for helpful feedback on the manuscript. This work is supported by the National Natural Science Foundation of China (grant Nos. 12273005, 12494575, and 12133003), the National Key R\&D Program of China (grant No. 2023YFE0117200 and 2024YFA1611700), the special funding for Guangxi Bagui Youth Scholars,
and the Guangxi Talent Program (``Highland of Innovation Talents'').
\end{acknowledgements}

\bibliography{sample631}

\end{document}